# Analysis of Inter-Testamental References Reveal Five Groups of Books in the Christian Bible

By Isaac Anderson, Wesley Stevick, and Dr. Katrina Koehler

## Abstract

The Bible is packed with references from start to finish. This study aims to analyze a specific branch of these references: citations. While there are several types of references, both explicit and implicit, this study focuses on the types of references that can be detected with a simple algorithmic string comparison, or an n-gram string comparison. Words were compared by their Strong's Concordance numbers so they could be compared without conjugation or declension. We searched through the Greek Old Testament (Septuagint) and Greek New Testament manuscripts for direct quotations from the former in the latter. Our analysis of these references leads us to believe Old Testament books cluster into three groups of common use, and that New Testament books cluster into two books of common use. We analyze these clusters to show explicitly how they differ, and discover that New Testament books reference vastly different portions of the Old Testament.

## Introduction

One of the fascinating features of the Bible is its rich-interconnectedness and the layers of reference and allegory carried by the text. Entire books are written on the Bible's literary use of itself. Some even consider this depth of textual self-reference to be evidence for the book's veracity: the Old

Testament is known to be written as early as the 10th century BCE, with most of the Torah being written in the 5th century BCE whereas most of the New Testament was written in the 1st and 2nd centuries CE.[1] If the forty or so individuals responsible for the Bible's creation were not inspired by the same God, it is argued, the work as a whole would not be such a self-consistent, self-referential narrative.

The work of Biblical interpretation is dominated by humans. This is good and necessary: the Bible is *for* humans, after all. Additionally, even the most powerful natural language processing techniques (currently LLMs) have yet to rival the best humans in nuanced, sophisticated understanding of text. Computational Bible study offers two advantages, however.

The first is objectivity: it is not possible for a human to read the Bible "fresh" - uninfluenced by everything they already believe. The Protestant doctrine of *sola scriptura* ("Only Scripture") is an impossible ideal, because a human cannot forget every extrabiblical belief that might influence their interpretation. Our work does not attain this ideal, as the questions we asked and the code we wrote were informed by knowledge we possess as members of a Christian culture; however the *answers* to those questions can more confidently be said to have come from the text itself, not our biases or preconceptions.

The second is clarity: human intuition far surpasses any current computational understanding; this is evident from the fact that the field of biblical scholarship is dominated by *natural* neural networks. However, the relative simplicity of numeric computation allows a level of exactitude not possible otherwise. We did not simply discover that the book of Revelation made "heavy" use of the Old Testament. We discovered that 12.2% of the words in Revelation were direct quotations (of at least five words in length) to books of the Old Testament, and that this value was higher for Revelation than any other book in the New Testament.

---

[1] The Editors of Encyclopædia Britannica, "When Was the Bible Written?," *Britannica*, accessed November 26, 2025, https://www.britannica.com/question/When-was-the-Bible-written.

There exist vast concordances of Biblical references and allusions of every kind, filled only by the vast levels of domain knowledge within its own field of theology. This said, there has been little work done in proposing a computational approach in examining direct quotations. Thus we propose a new methodology in examining Biblical subject matter and finding citations. Upon constructing this new tool, and conducting a general analysis of these references we then ask the following research questions: do the books of the Bible cluster well by referential use? If so, what are the characteristics of these clusters, and do they align with a modern theological understanding of the text?

## Methodology

### Understanding Citations and Allusions.

We recognise there to be two branches of Biblical references: Citations and Allusions/Verbal Parallels.[2] Citations can be made explicitly, or implicitly and can either paraphrase or quote verbatim. These, as one can imagine, are far easier to detect. Then we move onto the second type of Biblical reference, Allusions and Verbal Parallels. Allusions are by nature "shorter and subtler references" to an Old Testament text. There are many facets of them, such as "reminiscence" - "Have you not read what David did" (Luke 6:3-4) or specific allusions such as the "Son of Man" clearly being a reference to Daniel 7, but never explicitly mentioned so. Lastly, there are also thematic echoes, such as Jesus leading the people and feeding them in the gospels bearing similar imagery to Moses.[3]

Now that an understanding of Biblical allusions has been discussed, it is important to establish that the vast majority of Biblical references are Allusions and Verbal parallels and not Citations. For this exact reason it is important to understand that the proposed tool in this research paper; one that finds direct n-grams is not an entire or holistic representation of Biblical references, rather it is a much smaller subset. Thus, for the purpose of this paper "reference" refers only to a direct quotation, unless explicitly used otherwise.

---

[2] Huffman, Douglas S. "A Two-Dimensional Taxonomy of Forms for the NT Use of the OT." *Themelios* 46, no. 2 (2021): 306–18.

[3] BibleProject Scholarship Team, "The Sermon on the Mount and Jesus as the New Moses," *BibleProject*, August 15, 2018, https://bibleproject.com/blog/sermon-mount-jesus-new-moses/

## Quotation Detection

Initially the data was in XML format as a list of Greek words, each annotated with grammatical information and its Strong's number.[4] Each Strong's number corresponds to the unconjugated form of a word; by comparing the Strong's numbers of two words, it can be determined whether the underlying meaning is the same. To use an English example, "go," "going," and "gone" all count as the same word, but "departed" is distinct.

Some of the words did not come annotated with Strong's numbers. A new number (prefixed with "C-" for Custom, to distinguish from real Strong's numbers) was assigned to each distinct unlabeled word. These Custom Strong's numbers are not conjugation aware, however they do allow identical words in the Old and New Testaments to be flagged as such.

To detect matches, the following algorithm was performed: the Strong's numbers of each sequence of five consecutive Greek words (5-gram) from the Old Testament was compared with each 5-gram from the New. If a match occurred, the locations of the 5-grams were recorded for further processing. We did not search for 5-grams that spanned more than one book.

After this search, the database contained 6,388 locations into the Old and New Testaments where five consecutive words matched. However, this data contained duplications - where multiple 5-grams spanned consecutive verses. To fix this, all overlapping quotations were merged, producing 4,807 distinct quotations. See Figure 1 for the quotation length after this process.

---

[4]Jens Grabner, *The Analytic Septuagint*, XML ed. (Zefania XML Project, 2009), http://sourceforge.net/projects/zefania-sharp; and Ulrik Petersen, ed., *Tischendorf Greek New Testament (Strongs)*, XML ed. (Zefania XML Project, 2009), http://sourceforge.net/projects/zefania-sharp.

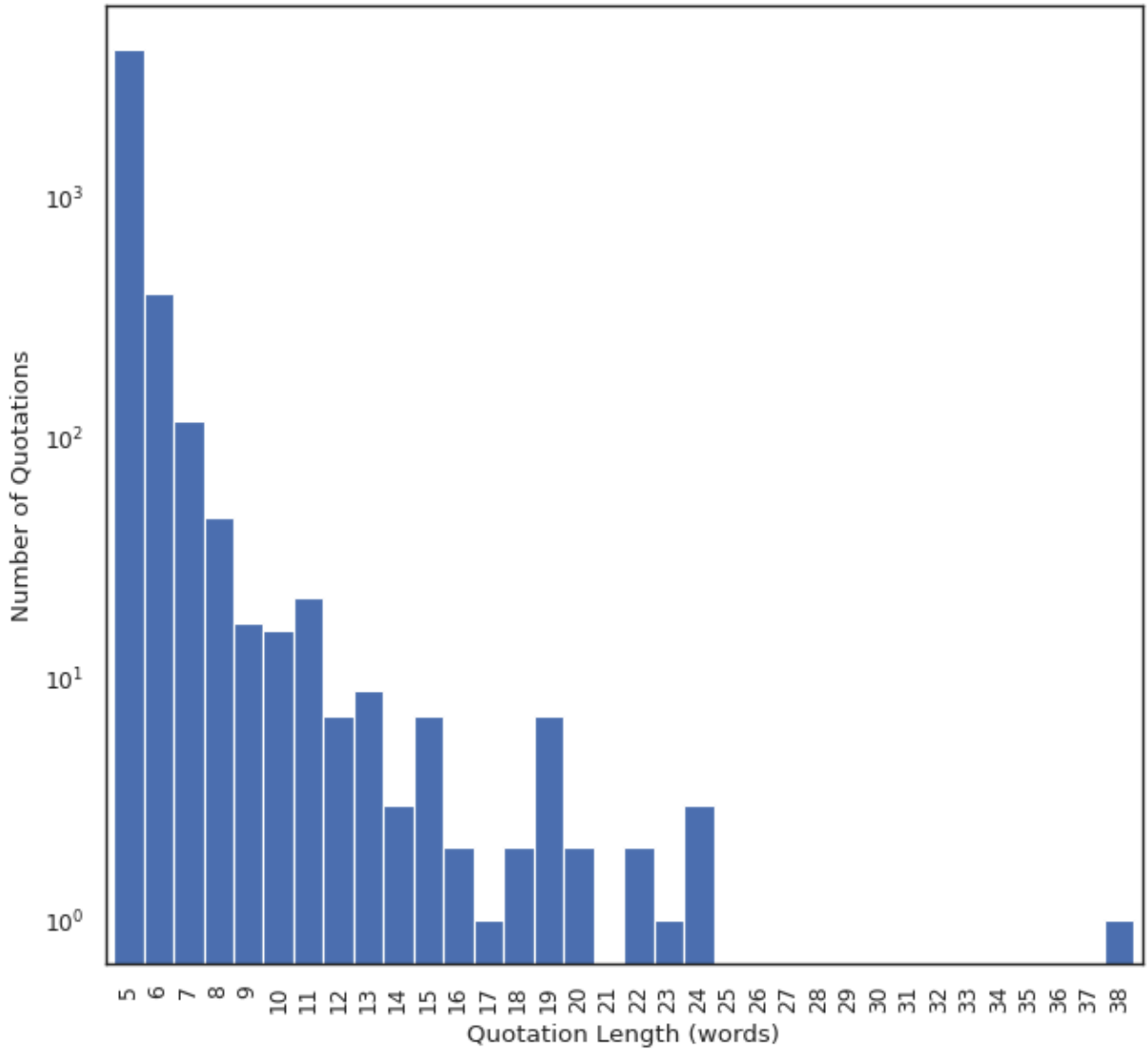

**Figure 1.** Distribution of lengths of quotations. Notice that most quotations are only 5-grams, with diminishingly few of greater length.

## Analysis

The final purpose of this study is to determine which New Testament books use the Old Testament in similar ways, and which Old Testament books are used together by the New Testament. First, a table was created with a row for each New Testament book and a column for each Old Testament book (see Appendix). The value of each cell is the proportion of words in the new book that are direct quotations from the old book. For instance, the 0.023 in the Revelation row under the Psalms column means that 2.3% of the words in Revelation are direct quotations from Psalms.

The logarithm of these proportions was then calculated (the logarithm of the smallest nonzero value was first added to each cell, to prevent zero results). The table produced can be interpreted either as a point in 39-dimensional space for each New Testament book, representing how it uses the Old Testament, or as a point in 27-dimensional space for each Old Testament book, representing how it is used in the New Testament. Hierarchical clustering was applied to both of these, using the Euclidean distance metric and ward linkage. This procedure was selected because it produced clean, relatively meaningful results.

## Literature Review

There are several works that we would like to discuss that have been both influential or parallel to our current research. These works will show that our research is in part unique.

One example of Biblical references is Chris Harrison's "Bible Cross References".[5] Before we proceed, a key difference between our study and Harrison's is that Harrison used a Biblical concordance which was peer reviewed by other scholars. That is to say, Harrison did not derive references using computational algorithms or Natural Processing. Where Harrison's study is pivotal however is an attempt to create beauty that also conveys the complexity of the Bible's cross referencing (Figure 1).

---

[5] Chris Harrison, "BibleViz (Bible Cross-References and Related Visualizations)," *chrisharrison.net*, accessed November 26, 2025, https://www.chrisharrison.net/index.php/Visualizations/BibleViz

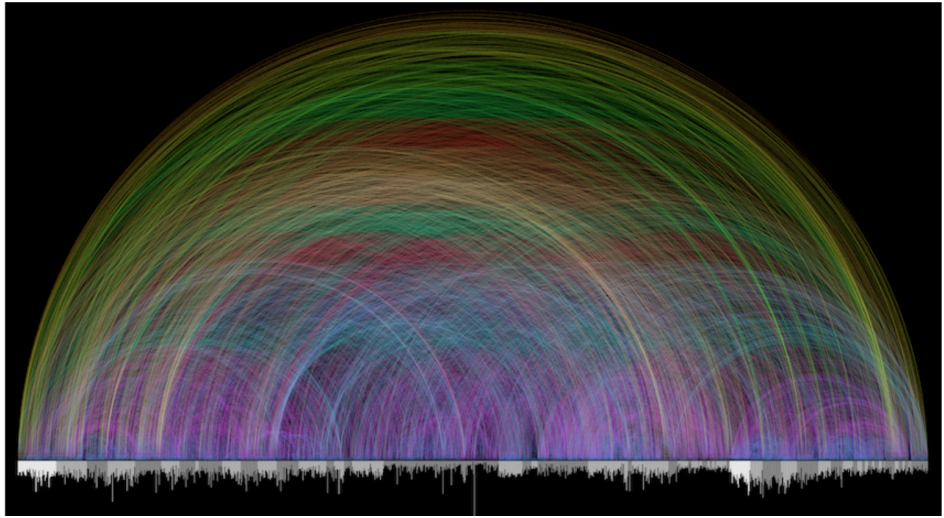

**Figure 1**. Each chapter of the Bible is visualized as a grey line, its length being the total word count, and then each book is coloured in a shade of grey that contours to the shades around it. Then Harrison maps out connections between chapters, with the colour of the connection being the distance.

To compare this work to our own we do note key differences. First and foremost, the use of a concordance but also the sheer size. The size of our reference database was 4,807; this is dwarfed by Harrison's 63,779 references. This leads to an interesting intersection of our previous acknowledgement that we can only detect citations and not allusions. If we are to take Harrison's concordance as the ultimate sort we see that our basic algorithm which detected citations only accounts for a minute percentage of the actual amount of references.

The second study which was influential to our own but again shows the difference in approach is "New in Labs: Cross References".[6] This study produced a "search engine" where you can search all the supposed references from one verse and find the others. The approach in this method was using TSK ("The Treasury of Scripture Knowledge") as well as Twitter posts and The Topical Bible. First and foremost, the TSK is a large concordance of such that as all citations. The Topical Bible is a search engine built from the Yahoo search engine and the ESV Bible that enables one to search for other verses based upon the topic one searches. Then lastly, the Twitter posts were dissected based upon what was being mentioned within the post and the verse that the post directly quoted itself. All together this meant the researchers were able to produce a system of weights and measures. This system of weights and measures, although not fully described, does change with user input, where users who search for Biblical cross references can also vote on whether the provided options are satisfactory or not. We find this to be an interesting solution as opposed to our more rigid approach of finding citations. This project did raise the question: how, if ever, can we discern what an allusion or reference is? Perhaps the most concrete way is to take the votes of many many readers and find what they agree on, as ultimately no amount of sophistication in algorithms will match the human mind.

---

[6] OpenBible.info, "New in Labs: Cross References," *OpenBible.info Blog*, April 11, 2010, https://www.openbible.info/blog/2010/04/new-in-labs-cross-references/

## Results

Clustering of the Old Testament produced three clusters, shown in Table 1. The clustering algorithm operates in 27-dimensional space, but a portion of the data can be seen in Figure 2. Likewise, see Table 2 for both New Testament, and Figure 3 for 2 of the 66 dimensions used to select these clusters.

| Book | Cluster | Book | Cluster |
|---|---|---|---|
| Genesis | OT3 | Ecclesiastes | OT2 |
| Exodus | OT1 | Song of Solomon | OT2 |
| Leviticus | OT1 | Isaiah | OT3 |
| Numbers | OT1 | Jeremiah | OT1 |
| Deuteronomy | OT1 | Lamentations | OT2 |
| Joshua | OT1 | Ezekiel | OT1 |
| Judges | OT1 | Daniel | OT1 |
| Ruth | OT2 | Hosea | OT2 |
| 1 Samuel | OT1 | Joel | OT2 |
| 2 Samuel | OT1 | Amos | OT2 |
| 1 Kings | OT1 | Obadiah | OT2 |
| 2 Kings | OT1 | Jonah | OT2 |
| 1 Chronicles | OT1 | Micah | OT2 |
| 2 Chronicles | OT1 | Nahum | OT2 |
| Ezra | OT2 | Habakkuk | OT2 |
| Nehemiah | OT1 | Zephaniah | OT2 |
| Esther | OT2 | Haggai | OT2 |
| Job | OT2 | Zechariah | OT1 |
| Psalms | OT3 | Malachi | OT2 |
| Proverbs | OT2 | | |

**Table 1**. The books contained in each of the three clusters. OT1 contains four books of the Torah, most of the history, and some prophets. OT2 contains the Wisdom literature and most of the prophets. OT3 contains the three most referenced books: Genesis, Psalms, and Isaiah.

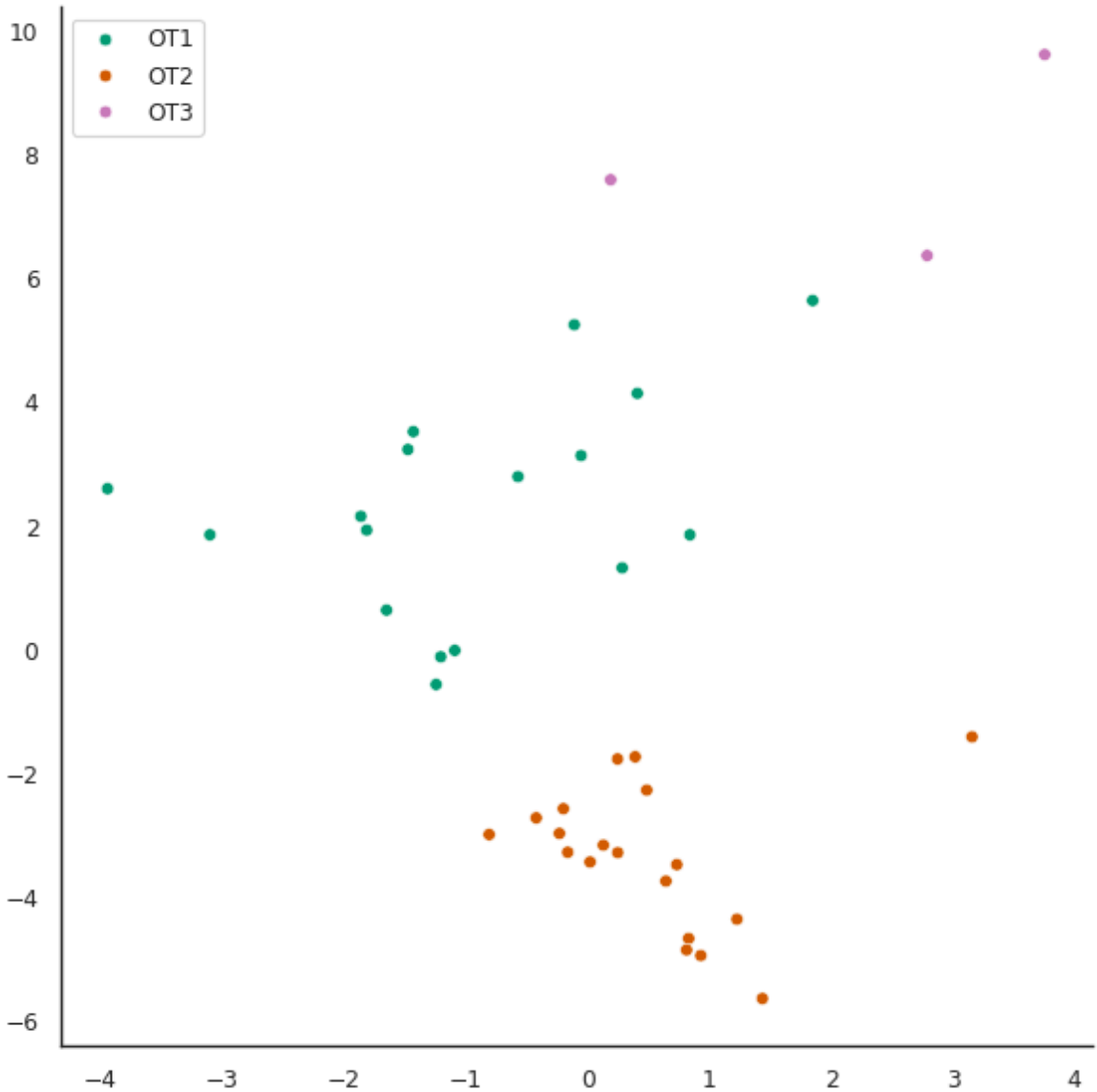

**Figure 2.** Each point represents one book of the Old Testament. The units are meaningless - x and y values were created by applying PCA to the log-proportion table as described in Analysis, explaining 67% of the variance. Observe that most of the Old Testament separates cleanly into OT1 and OT2, with only three books in OT3.

| Book | Cluster | Book | Cluster |
|---|---|---|---|
| Matthew | NT2 | 1 Timothy | NT1 |
| Mark | NT2 | 2 Timothy | NT1 |
| Luke | NT2 | Titus | NT1 |
| John | NT2 | Philemon | NT1 |
| Acts | NT2 | Hebrews | NT2 |
| Romans | NT1 | James | NT1 |
| 1 Corinthians | NT1 | 1 Peter | NT1 |
| 2 Corinthians | NT1 | 2 Peter | NT1 |
| Galatians | NT1 | 1 John | NT1 |
| Ephesians | NT1 | 2 John | NT1 |
| Philippians | NT1 | 3 John | NT1 |
| Colossians | NT1 | Jude | NT1 |
| 1 Thessalonians | NT1 | Revelation | NT2 |
| 2 Thessalonians | NT1 | | |

**Table 2.** Cluster 2 contains the Gospels (including Acts), Hebrews, and Revelation. These are the books with the highest reference-density, so it seems this is the "literary author" cluster.

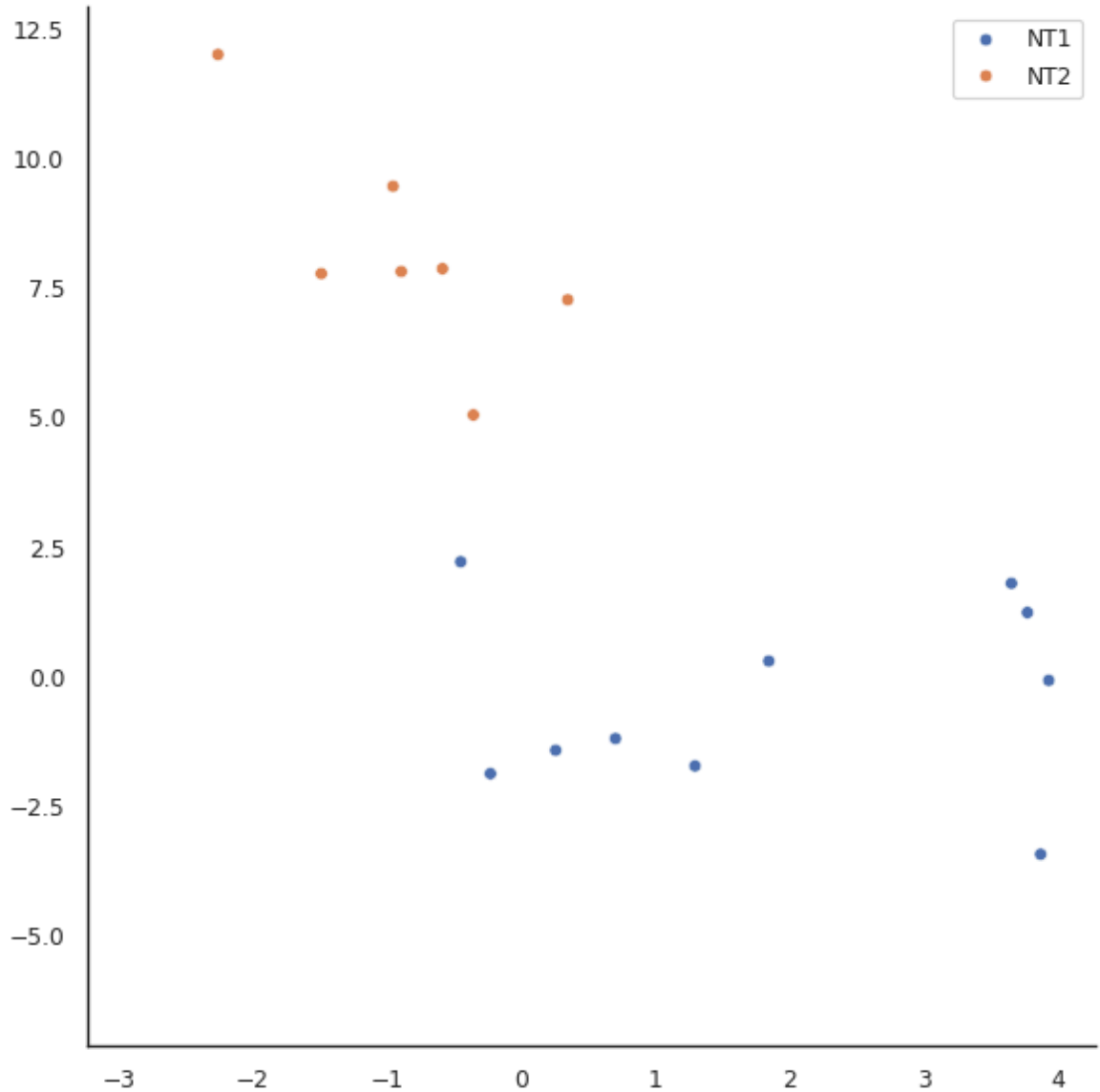

**Figure 3.** Each point corresponds to one book of the New Testament. The units are meaningless - x and y values were created by applying PCA to the log-proportion table described in Analysis, explaining 58% of the variance in the data.

## Statistics

See Table 3 for statistics on each cluster. The longest one by far is OT1 (including most of the Torah and historical texts), with roughly an order of magnitude more words than the others. The most reference-dense is NT2 (containing the Gospels, Hebrew, and Revelation), where more than one word in twenty is a direct quote from the Old Testament.

| Cluster | Number of books | Word Count | Total Reference Words | Mean Reference Length | Number of References | Quotation Density |
|---------|-----------------|------------|-----------------------|-----------------------|----------------------|-------------------|
| OT1 | 17 | 317173 | 9062 | 5.24 | 2884 | 2.86% |
| OT2 | 19 | 63622 | 1060 | 5.34 | 348 | 1.67% |
| OT3 | 3 | 94605 | 3660 | 5.5 | 1575 | 3.87% |
| NT1 | 20 | 39907 | 1017 | 5.51 | 464 | 2.55% |
| NT2 | 7 | 97618 | 6058 | 5.32 | 4343 | 6.21% |

**Table 3.** Reference words are either words quoted from the Old Testament (in a NT cluster), or words that will be quoted in the New Testament (in an OT cluster). Quotation density is Total Reference Words / Word Count: the proportion of words in the cluster that are involved in a quotation.

## Inter-use of Clusters

Figure 4 shows how many of the references in NT1 and NT2 are to each of the Old Testament clusters. Figure 5 shows how many references in OT1, OT2, and OT3 are by each of the New Testament clusters. For example: we see that 2,726 of the references in NT2 are to OT1.

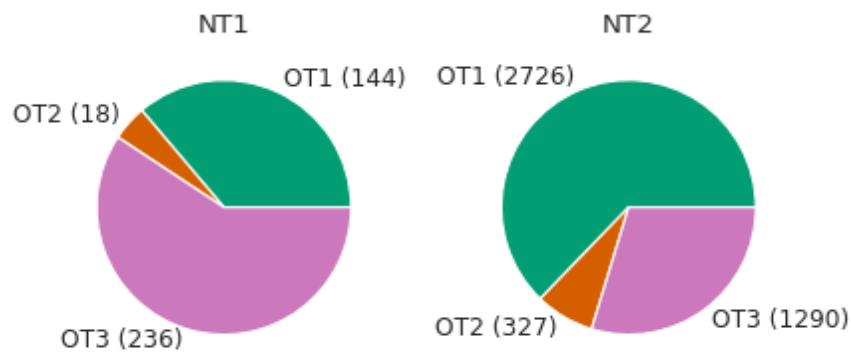

**Figure 4.** Numbers in parentheses are reference counts. NT1 (non-Gospel cluster) uses mostly OT3 (Genesis/Isaiah/Psalms), whereas NT2 uses mostly OT1 (Torah + History).

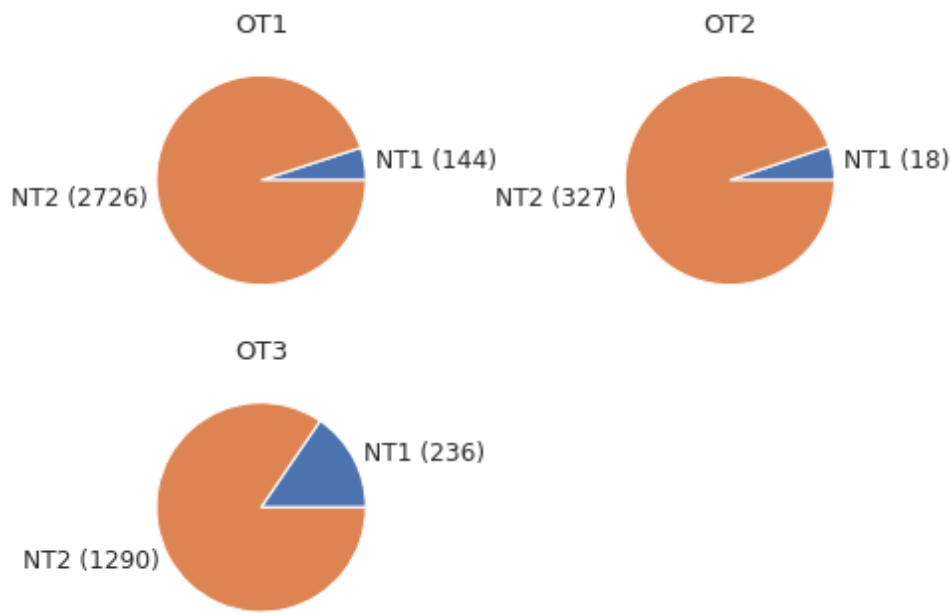

**Figure 5.** Numbers in parentheses are reference counts. Notice most references are from NT2, the "Gospels" cluster. Also notice that OT3 (Genesis/Isaiah/Psalms) has many more NT1 references.

## Discussion

The clusters generated for the Old Testament appear meaningful. The most referenced "heavy hitters" (Genesis, Psalms, Isaiah) form one cluster. OT1 contains the rest of the Torah, as well as a substantial amount of the history (Samuel, Kings, and Chronicles), as well as a few prophets. OT2 contains the Wisdom literature (Song of Solomon, Job, Proverbs, Ecclesiastes, Lamentations), and more prophets. These may not be meaningful, as the clustering method used was selected from others because it produced interpretable clusters. Further research will need to be done, however it appears that the ways New Testament authors used Old Testament books align at least partially with the groupings applied to the Old Testament.

The clusters generated for the New Testament plainly are meaningful. It is likely not coincidental that the gospels form a cluster together, that John stands off a little from the synoptics, or that Acts was classified with Luke. Further research must be done to determine the relevance of the inclusion of Hebrews and Revelation in this category.

## Conclusion

Our research shows the Bible to contain five referentially similar collections of books - three in the Old Testament, and two in the New. This separation was done computationally, relying exclusively on information stored in the text itself, rather than a human linguistic understanding. Despite the limited information available to the algorithm, it produced labels that mostly aligned with a human understanding of the text: "Torah and historical texts," "Gospels and Hebrew-literate texts," "important Old Testament book," etc. We were able to uncover information about the most quoted (thus arguably most important) sections of the Bible, but also information about which sections of the Bible rely on which: that the Gospels quote more Torah and historical texts, while other New Testament books mostly quote Genesis, Psalms, and Isaiah. It is now the job of a theologian to tell us what this means.

### Future work

Our method for detecting references is primitive. The current analysis doesn't understand synonyms, paraphrasing, allusions of less than 5 words (such as "son of man"), or more complex structural parallels. It is likely that significant Biblical connections were ignored because they were too complicated to detect.

Additionally, it must be noted that the parameters used to create these clusters were selected *because they produced interpretable clusters*. It would be valuable to apply this same method to another, similar text, to see if it consistently detects semantic meaning across corpora.

Further research should also implement a method of quantifying the separation between clusters, comparing ours to the commonly ascribed labels (TaNaK, Gospels/Epistles, etc.).

# Appendix: Table of quotation proportions

## Genesis - Deuteronomy

|  | Genesis | Exodus | Leviticus | Numbers | Deuteronomy |
|---|---|---|---|---|---|
| Matthew | 1.40% | 0.41% | 0.15% | 0.51% | 0.84% |
| Mark | 1.17% | 0.56% | 0.22% | 0.50% | 0.86% |
| Luke | 1.47% | 0.79% | 0.27% | 0.51% | 1.11% |
| John | 1.03% | 0.17% | 0.10% | 0.10% | 0.24% |
| Acts | 0.86% | 0.94% | 0.25% | 0.37% | 0.58% |
| Romans | 0.82% | 0.41% | 0.07% | 0.00% | 0.51% |
| 1 Corinthians | 0.07% | 0.22% | 0.07% | 0.00% | 0.07% |
| 2 Corinthians | 0.11% | 0.11% | 0.00% | 0.22% | 0.34% |
| Galatians | 1.44% | 0.22% | 0.00% | 0.00% | 0.31% |
| Ephesians | 0.79% | 0.70% | 0.00% | 0.21% | 0.70% |
| Philippians | 0.00% | 0.31% | 0.00% | 0.00% | 0.00% |
| Colossians | 0.00% | 0.00% | 0.00% | 0.00% | 0.00% |
| 1 Thessalonians | 0.00% | 0.68% | 0.00% | 0.00% | 0.00% |
| 2 Thessalonians | 1.22% | 0.00% | 0.00% | 0.00% | 0.00% |
| 1 Timothy | 0.00% | 0.00% | 0.00% | 0.00% | 0.00% |
| 2 Timothy | 0.00% | 0.00% | 0.00% | 0.40% | 0.40% |
| Titus | 0.00% | 0.00% | 0.00% | 0.00% | 0.00% |
| Philemon | 0.00% | 0.00% | 0.00% | 0.00% | 0.00% |
| Hebrews | 0.95% | 1.13% | 0.30% | 0.61% | 0.24% |
| James | 0.40% | 0.29% | 0.00% | 0.00% | 0.29% |
| 1 Peter | 0.00% | 0.00% | 0.00% | 0.00% | 0.36% |
| 2 Peter | 0.00% | 0.00% | 0.00% | 0.00% | 0.00% |
| 1 John | 1.87% | 0.23% | 0.28% | 0.93% | 0.47% |
| 2 John | 0.00% | 0.00% | 0.00% | 0.00% | 0.00% |
| 3 John | 0.00% | 0.00% | 0.00% | 0.00% | 0.00% |
| Jude | 0.00% | 1.08% | 0.00% | 0.00% | 0.00% |
| Revelation | 1.67% | 1.73% | 0.49% | 0.73% | 1.17% |

## Joshua - 2 Samuel

|  | Joshua | Judges | Ruth | 1 Samuel | 2 Samuel |
|---|---|---|---|---|---|
| Matthew | 0.70% | 0.54% | 0.14% | 0.42% | 0.50% |
| Mark | 0.89% | 0.34% | 0.19% | 0.27% | 0.51% |
| Luke | 0.73% | 0.75% | 0.19% | 0.96% | 0.89% |
| John | 0.56% | 0.26% | 0.06% | 0.36% | 0.29% |
| Acts | 0.35% | 0.30% | 0.06% | 0.58% | 0.51% |
| Romans | 0.28% | 0.14% | 0.00% | 0.07% | 0.00% |
| 1 Corinthians | 0.07% | 0.15% | 0.00% | 0.00% | 0.07% |
| 2 Corinthians | 0.00% | 0.00% | 0.00% | 0.11% | 0.11% |
| Galatians | 0.36% | 0.22% | 0.00% | 0.00% | 0.00% |
| Ephesians | 0.21% | 0.41% | 0.25% | 0.00% | 0.21% |
| Philippians | 0.00% | 0.00% | 0.00% | 0.00% | 0.00% |
| Colossians | 0.00% | 0.00% | 0.00% | 0.00% | 0.00% |
| 1 Thessalonians | 0.00% | 0.00% | 0.00% | 0.00% | 0.00% |
| 2 Thessalonians | 0.00% | 0.00% | 0.00% | 0.00% | 0.61% |
| 1 Timothy | 0.00% | 0.00% | 0.00% | 0.00% | 0.00% |
| 2 Timothy | 0.00% | 0.00% | 0.00% | 0.00% | 0.40% |
| Titus | 0.00% | 0.00% | 0.00% | 0.00% | 0.00% |
| Philemon | 0.00% | 0.00% | 0.00% | 0.00% | 0.00% |
| Hebrews | 0.22% | 0.30% | 0.00% | 0.20% | 0.42% |
| James | 0.00% | 0.00% | 0.00% | 0.29% | 0.29% |
| 1 Peter | 0.00% | 0.00% | 0.00% | 0.00% | 0.00% |
| 2 Peter | 0.00% | 0.00% | 0.00% | 0.00% | 0.00% |
| 1 John | 0.47% | 0.00% | 0.00% | 0.23% | 0.23% |
| 2 John | 0.00% | 0.00% | 0.00% | 0.00% | 0.00% |
| 3 John | 0.00% | 0.00% | 0.00% | 0.00% | 0.00% |
| Jude | 0.00% | 0.00% | 0.00% | 0.00% | 0.00% |
| Revelation | 0.87% | 0.48% | 0.10% | 1.16% | 0.62% |

## 1 Kings - Ezra

|  | 1 Kings | 2 Kings | 1 Chronicles | 2 Chronicles | Ezra |
|---|---|---|---|---|---|
| Matthew | 0.52% | 0.50% | 0.29% | 0.31% | 0.08% |
| Mark | 0.46% | 0.47% | 0.25% | 0.73% | 0.14% |
| Luke | 1.00% | 0.67% | 0.34% | 0.93% | 0.16% |
| John | 0.17% | 0.28% | 0.18% | 0.46% | 0.08% |
| Acts | 0.54% | 0.72% | 0.36% | 0.32% | 0.14% |
| Romans | 0.24% | 0.07% | 0.00% | 0.14% | 0.00% |
| 1 Corinthians | 0.00% | 0.00% | 0.00% | 0.07% | 0.00% |
| 2 Corinthians | 0.00% | 0.00% | 0.11% | 0.00% | 0.00% |
| Galatians | 0.00% | 0.00% | 0.00% | 0.31% | 0.00% |
| Ephesians | 0.00% | 0.00% | 0.00% | 0.21% | 0.00% |
| Philippians | 0.31% | 0.00% | 0.00% | 0.00% | 0.37% |
| Colossians | 0.00% | 0.00% | 1.02% | 0.00% | 0.00% |
| 1 Thessalonians | 0.00% | 0.00% | 0.34% | 0.00% | 0.00% |
| 2 Thessalonians | 0.61% | 0.00% | 0.00% | 0.00% | 0.00% |
| 1 Timothy | 0.00% | 0.00% | 0.00% | 0.31% | 0.00% |
| 2 Timothy | 0.00% | 0.00% | 0.81% | 0.89% | 0.00% |
| Titus | 0.00% | 0.00% | 0.00% | 0.00% | 0.00% |
| Philemon | 0.00% | 0.00% | 0.00% | 0.00% | 0.00% |
| Hebrews | 0.55% | 0.22% | 0.53% | 0.42% | 0.00% |
| James | 0.58% | 0.00% | 0.29% | 0.29% | 0.00% |
| 1 Peter | 0.00% | 0.00% | 0.00% | 0.00% | 0.00% |
| 2 Peter | 0.00% | 0.00% | 0.00% | 0.00% | 0.00% |
| 1 John | 0.47% | 0.00% | 0.23% | 0.23% | 0.00% |
| 2 John | 0.00% | 0.00% | 0.00% | 0.00% | 0.00% |
| 3 John | 0.00% | 0.00% | 0.00% | 0.00% | 0.00% |
| Jude | 0.00% | 0.00% | 0.00% | 0.00% | 0.00% |
| Revelation | 0.82% | 1.48% | 0.84% | 1.60% | 0.19% |

## Nehemiah - Proverbs

| | Nehemiah | Esther | Job | Psalms | Proverbs |
|---|---|---|---|---|---|
| Matthew | 0.22% | 0.08% | 0.05% | 0.93% | 0.08% |
| Mark | 0.36% | 0.23% | 0.09% | 1.16% | 0.00% |
| Luke | 0.19% | 0.11% | 0.00% | 0.97% | 0.06% |
| John | 0.26% | 0.10% | 0.14% | 0.62% | 0.03% |
| Acts | 0.29% | 0.11% | 0.11% | 1.10% | 0.03% |
| Romans | 0.00% | 0.00% | 0.00% | 2.04% | 0.38% |
| 1 Corinthians | 0.07% | 0.00% | 0.00% | 0.15% | 0.00% |
| 2 Corinthians | 0.00% | 0.00% | 0.00% | 0.38% | 0.00% |
| Galatians | 0.00% | 0.22% | 0.00% | 0.67% | 0.00% |
| Ephesians | 0.00% | 0.00% | 0.00% | 0.62% | 0.21% |
| Philippians | 0.00% | 0.00% | 0.31% | 0.31% | 0.00% |
| Colossians | 0.00% | 0.00% | 0.00% | 0.00% | 0.00% |
| 1 Thessalonians | 0.00% | 0.00% | 0.00% | 0.00% | 0.00% |
| 2 Thessalonians | 0.00% | 0.00% | 0.00% | 0.61% | 0.00% |
| 1 Timothy | 0.00% | 0.00% | 0.00% | 0.31% | 0.00% |
| 2 Timothy | 0.00% | 0.40% | 0.00% | 0.81% | 0.00% |
| Titus | 0.00% | 0.00% | 0.00% | 0.00% | 0.00% |
| Philemon | 0.00% | 0.00% | 0.00% | 0.00% | 0.00% |
| Hebrews | 0.22% | 0.00% | 0.00% | 5.54% | 0.38% |
| James | 0.00% | 0.00% | 0.00% | 0.58% | 0.00% |
| 1 Peter | 0.00% | 0.00% | 0.00% | 3.52% | 0.42% |
| 2 Peter | 0.00% | 0.00% | 0.00% | 0.45% | 0.00% |
| 1 John | 0.00% | 0.00% | 0.00% | 0.70% | 0.00% |
| 2 John | 0.00% | 0.00% | 0.00% | 0.00% | 0.00% |
| 3 John | 0.00% | 0.00% | 0.00% | 0.00% | 0.00% |
| Jude | 0.00% | 0.00% | 0.00% | 0.00% | 0.00% |
| Revelation | 0.69% | 0.25% | 0.20% | 2.30% | 0.15% |

## Ecclesiastes - Lamentations

| | Ecclesiastes | Song of Solomon | Isaiah | Jeremiah | Lamentations |
|---|---|---|---|---|---|
| Matthew | 0.08% | 0.00% | 0.91% | 0.47% | 0.00% |
| Mark | 0.09% | 0.00% | 0.68% | 0.41% | 0.00% |
| Luke | 0.06% | 0.00% | 0.95% | 0.85% | 0.03% |
| John | 0.03% | 0.00% | 0.19% | 0.37% | 0.00% |
| Acts | 0.03% | 0.00% | 0.98% | 0.48% | 0.03% |
| Romans | 0.00% | 0.00% | 1.13% | 0.00% | 0.00% |

| 1 Corinthians | 0.00% | 0.00% | 0.28% | 0.15% | 0.00% |
| 2 Corinthians | 0.00% | 0.00% | 0.31% | 0.11% | 0.00% |
| Galatians | 0.00% | 0.00% | 0.76% | 0.31% | 0.00% |
| Ephesians | 0.00% | 0.00% | 0.21% | 0.21% | 0.00% |
| Philippians | 0.00% | 0.00% | 0.00% | 0.00% | 0.00% |
| Colossians | 0.00% | 0.00% | 0.32% | 0.00% | 0.00% |
| 1 Thessalonians | 0.00% | 0.00% | 0.00% | 0.00% | 0.00% |
| 2 Thessalonians | 0.00% | 0.00% | 1.09% | 0.00% | 0.00% |
| 1 Timothy | 0.00% | 0.00% | 0.00% | 0.00% | 0.00% |
| 2 Timothy | 0.00% | 0.00% | 0.00% | 0.00% | 0.00% |
| Titus | 0.00% | 0.00% | 0.00% | 0.00% | 0.00% |
| Philemon | 0.00% | 0.00% | 0.00% | 0.00% | 0.00% |
| Hebrews | 0.00% | 0.00% | 0.30% | 2.34% | 0.00% |
| James | 0.00% | 0.00% | 0.92% | 0.00% | 0.00% |
| 1 Peter | 0.00% | 0.00% | 1.49% | 0.00% | 0.00% |
| 2 Peter | 0.00% | 0.00% | 0.00% | 0.00% | 0.00% |
| 1 John | 0.23% | 0.00% | 0.00% | 0.00% | 0.00% |
| 2 John | 0.00% | 0.00% | 0.00% | 0.00% | 0.00% |
| 3 John | 0.00% | 0.00% | 0.00% | 0.00% | 0.00% |
| Jude | 0.00% | 0.00% | 0.00% | 0.00% | 0.00% |
| Revelation | 0.05% | 0.00% | 1.47% | 1.50% | 0.05% |

Ezekiel - Amos

| Matthew | 0.48% | 0.33% | 0.14% | 0.09% | 0.00% |
| Mark | 0.35% | 0.22% | 0.04% | 0.05% | 0.04% |
| Luke | 0.73% | 0.39% | 0.16% | 0.11% | 0.05% |
| John | 0.10% | 0.26% | 0.03% | 0.00% | 0.03% |
| Acts | 0.41% | 0.20% | 0.11% | 0.33% | 0.22% |
| Romans | 0.14% | 0.14% | 0.23% | 0.00% | 0.00% |
| 1 Corinthians | 0.00% | 0.00% | 0.00% | 0.00% | 0.00% |
| 2 Corinthians | 0.25% | 0.00% | 0.00% | 0.00% | 0.00% |
| Galatians | 0.00% | 0.00% | 0.00% | 0.00% | 0.00% |
| Ephesians | 0.00% | 0.21% | 0.00% | 0.00% | 0.00% |
| Philippians | 0.00% | 0.37% | 0.00% | 0.00% | 0.00% |
| Colossians | 0.32% | 0.00% | 0.00% | 0.38% | 0.00% |
| 1 Thessalonians | 0.34% | 0.00% | 0.00% | 0.00% | 0.00% |
| 2 Thessalonians | 0.61% | 0.00% | 0.00% | 0.00% | 0.00% |
| 1 Timothy | 0.00% | 0.00% | 0.00% | 0.00% | 0.00% |
| 2 Timothy | 0.00% | 0.00% | 0.00% | 0.00% | 0.00% |
| Titus | 0.00% | 0.00% | 0.00% | 0.00% | 0.00% |
| Philemon | 0.00% | 0.00% | 0.00% | 0.00% | 0.00% |
| Hebrews | 0.44% | 0.20% | 0.10% | 0.00% | 0.10% |
| James | 0.00% | 0.40% | 0.00% | 0.00% | 0.00% |
| 1 Peter | 0.00% | 0.00% | 0.00% | 0.00% | 0.00% |
| 2 Peter | 0.00% | 0.00% | 0.00% | 0.00% | 0.00% |
| 1 John | 0.23% | 0.00% | 0.00% | 0.00% | 0.00% |
| 2 John | 2.03% | 0.00% | 0.00% | 0.00% | 0.00% |
| 3 John | 0.00% | 0.00% | 0.00% | 0.00% | 0.00% |
| Jude | 0.00% | 0.00% | 0.00% | 0.00% | 0.00% |
| Revelation | 1.99% | 1.06% | 0.41% | 0.22% | 0.40% |

Obadiah - Habakkuk

| | Obadiah | Jonah | Micah | Nahum | Habakkuk |
|---|---|---|---|---|---|
| Matthew | 0.00% | 0.12% | 0.05% | 0.00% | 0.03% |
| Mark | 0.04% | 0.09% | 0.09% | 0.00% | 0.00% |
| Luke | 0.03% | 0.03% | 0.11% | 0.00% | 0.03% |
| John | 0.03% | 0.03% | 0.03% | 0.00% | 0.00% |
| Acts | 0.00% | 0.03% | 0.08% | 0.00% | 0.00% |
| Romans | 0.00% | 0.00% | 0.00% | 0.00% | 0.07% |
| 1 Corinthians | 0.00% | 0.00% | 0.00% | 0.00% | 0.00% |
| 2 Corinthians | 0.00% | 0.00% | 0.00% | 0.00% | 0.00% |
| Galatians | 0.00% | 0.00% | 0.00% | 0.00% | 0.00% |
| Ephesians | 0.00% | 0.00% | 0.00% | 0.00% | 0.00% |
| Philippians | 0.00% | 0.00% | 0.00% | 0.00% | 0.00% |
| Colossians | 0.00% | 0.00% | 0.00% | 0.00% | 0.00% |
| 1 Thessalonians | 0.00% | 0.00% | 0.00% | 0.00% | 0.00% |
| 2 Thessalonians | 0.00% | 0.00% | 0.00% | 0.00% | 0.00% |
| 1 Timothy | 0.00% | 0.00% | 0.00% | 0.00% | 0.00% |
| 2 Timothy | 0.00% | 0.00% | 0.00% | 0.00% | 0.00% |
| Titus | 0.00% | 0.00% | 0.00% | 0.00% | 0.00% |
| Philemon | 0.00% | 0.00% | 0.00% | 0.00% | 0.00% |
| Hebrews | 0.00% | 0.00% | 0.20% | 0.00% | 0.12% |
| James | 0.00% | 0.00% | 0.00% | 0.00% | 0.00% |
| 1 Peter | 0.00% | 0.00% | 0.00% | 0.00% | 0.00% |
| 2 Peter | 0.00% | 0.00% | 0.00% | 0.00% | 0.00% |
| 1 John | 0.00% | 0.00% | 0.00% | 0.00% | 0.23% |
| 2 John | 0.00% | 0.00% | 0.00% | 0.00% | 0.00% |
| 3 John | 0.00% | 0.00% | 0.00% | 0.00% | 0.00% |
| Jude | 0.00% | 0.00% | 0.00% | 0.00% | 0.00% |
| Revelation | 0.05% | 0.33% | 0.00% | 0.41% | 0.05% |

Zephaniah - Malachi

| | Zephaniah | Haggai | Zechariah | Malachi |
|---|---|---|---|---|
| Matthew | 0.05% | 0.05% | 0.22% | 0.00% |
| Mark | 0.04% | 0.00% | 0.28% | 0.00% |
| Luke | 0.10% | 0.03% | 0.41% | 0.08% |
| John | 0.06% | 0.03% | 0.06% | 0.00% |
| Acts | 0.03% | 0.05% | 0.09% | 0.04% |
| Romans | 0.00% | 0.00% | 0.00% | 0.07% |
| 1 Corinthians | 0.00% | 0.00% | 0.07% | 0.00% |
| 2 Corinthians | 0.00% | 0.00% | 0.00% | 0.00% |
| Galatians | 0.00% | 0.00% | 0.00% | 0.00% |
| Ephesians | 0.00% | 0.00% | 0.00% | 0.00% |
| Philippians | 0.00% | 0.00% | 0.00% | 0.00% |
| Colossians | 0.00% | 0.00% | 0.00% | 0.00% |
| 1 Thessalonians | 0.00% | 0.00% | 0.00% | 0.00% |
| 2 Thessalonians | 0.00% | 0.00% | 0.61% | 0.00% |
| 1 Timothy | 0.00% | 0.00% | 0.00% | 0.00% |
| 2 Timothy | 0.40% | 0.00% | 0.00% | 0.00% |
| Titus | 0.00% | 0.00% | 0.00% | 0.00% |
| Philemon | 0.00% | 0.00% | 0.00% | 0.00% |
| Hebrews | 0.00% | 0.20% | 0.40% | 0.00% |
| James | 0.00% | 0.00% | 0.00% | 0.00% |
| 1 Peter | 0.00% | 0.00% | 0.00% | 0.00% |
| 2 Peter | 0.00% | 0.00% | 0.00% | 0.00% |
| 1 John | 0.00% | 0.47% | 0.00% | 0.00% |
| 2 John | 0.00% | 0.00% | 0.00% | 0.00% |
| 3 John | 0.00% | 0.00% | 0.00% | 0.00% |
| Jude | 0.00% | 0.00% | 0.00% | 0.00% |
| Revelation | 0.00% | 0.15% | 0.71% | 0.05% |